# CARTEpigenoQC: A Quality Control Toolkit for CAR-T Single-Cell Epigenomic Data

Kaitao Lai[1]

2025-07-28

[1] University of Sydney

## 1 Summary

CARTEpigenoQC is an R-based toolkit designed to streamline quality control (QC) for single-cell epigenomic datasets involving Chimeric Antigen Receptor (CAR)-engineered T cells. With the growing application of scATAC-seq, scCUT&Tag, and scBS-seq to characterize CAR-T cell states (Satpathy et al. 2019), it has become critical to perform customized QC that not only addresses standard metrics like FRiP (Fraction of Reads in Peaks) and TSS enrichment, but also directly detects signal from CAR vector insertion sites.

CARTEpigenoQC supports both 10x Genomics and non-10x data formats and produces HTML and PNG summary outputs suited for exploratory analysis and regulatory-grade preclinical reporting. It is intended to assist researchers, core facilities, and translational immunologists in ensuring the validity of single-cell epigenomic profiling of engineered T cells (Finck and al. 2022; Robinson, Nguyen, and Sadelain 2023).

## 2 Statement of Need

Despite the rise in single-cell chromatin profiling technologies in immunotherapy, current QC pipelines do not provide functionality to assess CAR vector integration in scATAC-seq or related datasets. Existing tools like ArchR (Granja et al. 2021), Signac (Stuart et al. 2021), or SnapATAC (Fang et al. 2021) focus on general chromatin metrics but lack transgene-specific support. CARTEpigenoQC fills this gap by integrating standard QC (e.g., FRiP, duplication, peak overlap) with direct inspection of vector signal such as EF1-$\alpha$, PGK1, or TRAC locus insertions that are common in CAR constructs (Eyquem et al. 2017).

This tool was inspired by the increasing use of vector-based gene therapies and the demand for preclinical quality pipelines (Maude et al. 2018). It is lightweight, fast, and does not require complex dependencies. It generates clean, reproducible Markdown-based reports suitable for clinical handoff, regulatory review, or publication.

## 3 Features

- FRiP calculation per cell using fragment overlaps with peak regions.
- Visualization of CAR vector site coverage across single cells.
- Markdown-based report with interactive and static plots.
- Compatible with 10x Genomics 'fragments.tsv.gz', ArchR fragments, and BED-formatted CAR insertion sites.
- Designed for reproducibility and integration into preclinical immunotherapy pipelines.

# 4 Example Usage

CARTEpigenoQC includes a 'run_qc.R' script and report template. When supplied with fragment files and a BED file of CAR insertion coordinates, it calculates QC metrics and generates a self-contained HTML report.

# 5 Example Outputs

## 5.1 FRiP Score Distribution

This histogram visualizes the distribution of FRiP (Fraction of Reads in Peaks) across all cells. A peak around 0.2–0.4 is expected for high-quality scATAC-seq data (Chen et al. 2019). A right-skewed distribution is typical, and cells with FRiP < 0.2 may be filtered out in downstream analyses.

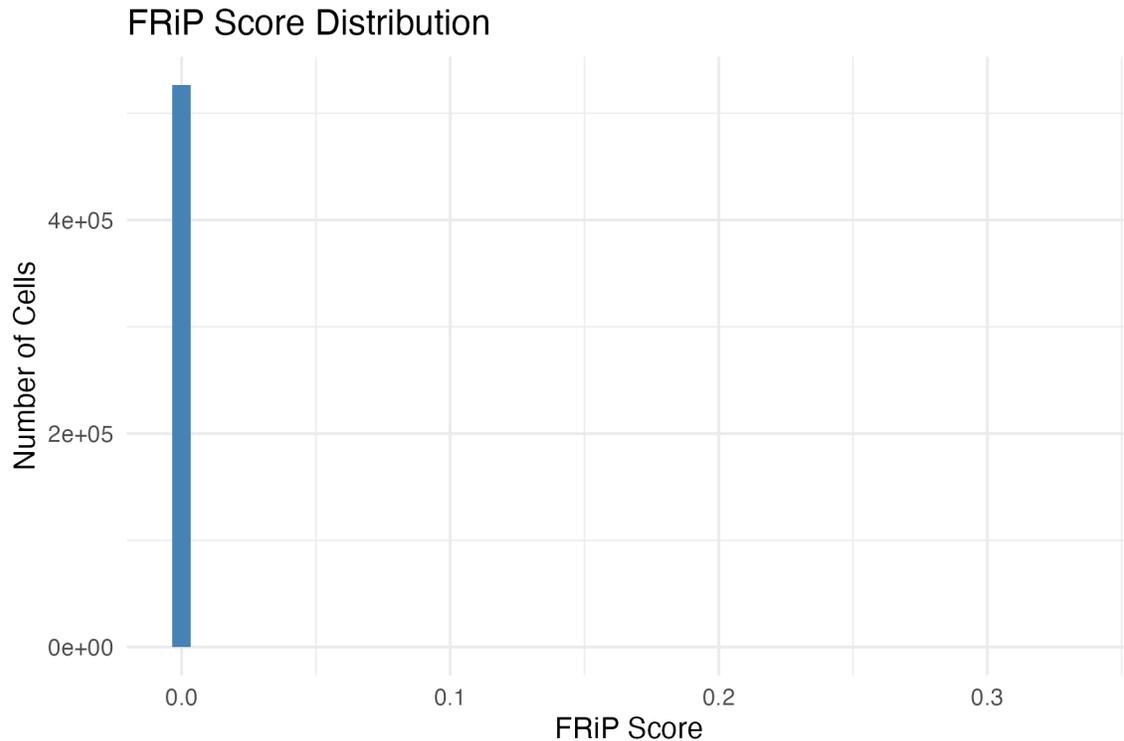

Figure 1: FRiP Histogram: Distribution of FRiP (Fraction of Reads in Peaks) scores across all single cells. This histogram visualizes the signal enrichment, with higher FRiP values indicating better signal-to-noise ratios and higher data quality in single-cell chromatin profiling.

## 5.2 Top Cells by FRiP

This table lists the top 20 cells with the highest FRiP scores. These cells often exhibit clearer chromatin signatures and are more informative for downstream applications like motif enrichment or regulatory network analysis.

## 5.3 Top FRiP Table

| # | Barcode | Total | In Peaks | FRiP |
|---|---|---|---|---|
| 1 | ACAGAAACAGGTCTGC-1 | 3 | 1 | 0.333333333333333 |
| 2 | ACTGTCCCATAGACGG-1 | 6 | 1 | 0.166666666666667 |

| #  | Barcode              | Total | In Peaks | FRiP              |
|----|----------------------|-------|----------|-------------------|
| 3  | GCAGCTGTCAATTGGC-1   | 17    | 1        | 0.058823529411765 |
| 4  | TTGTCTAAGCAGTAGC-1   | 24    | 1        | 0.041666666666667 |
| 5  | AGTTACGGTATCTGCA-1   | 1457  | 2        | 0.001372683596431 |
| 6  | CAGTATGAGTCGTATC-1   | 781   | 1        | 0.001280409731114 |
| 7  | ACTATTCTCGGATAAA-1   | 1695  | 2        | 0.001179941002950 |
| 8  | TCAGTCCGTGGACGAT-1   | 957   | 1        | 0.001044932079415 |
| 9  | AATGCCAGTGGCCTCA-1   | 966   | 1        | 0.001035196687371 |
| 10 | AATGGCTGTCCCTTTG-1   | 1002  | 1        | 0.000998003992016 |
| 11 | GCAGATTAGCCCATGC-1   | 1009  | 1        | 0.000991080277502 |
| 12 | AGTCAACAGGCGTCCT-1   | 1020  | 1        | 0.000980392156863 |
| 13 | TGCCTGTTCGCAGATT-1   | 1021  | 1        | 0.000979431929481 |
| 14 | CCCTCTCTCTTAACGG-1   | 1059  | 1        | 0.000944287063267 |
| 15 | CACATGACATATTGGC-1   | 1064  | 1        | 0.000939849624060 |
| 16 | CGCTATCTCGCTCTAC-1   | 1094  | 1        | 0.000914076782450 |
| 17 | TTAACGGAGTCGATAA-1   | 2228  | 2        | 0.000897666068223 |
| 18 | GAAGAGCCAAAGCATA-1   | 1208  | 1        | 0.000827814569536 |
| 19 | CCCGTTAAGGCAATTA-1   | 1234  | 1        | 0.000810372771475 |
| 20 | CAACGTAGTAAGTCTC-1   | 1243  | 1        | 0.000804505229284 |

Table 1: Top FRiP: Top 20 barcodes ranked by FRiP (Fraction of Reads in Peaks) scores. This table summarizes the total number of reads, number of reads overlapping peaks, and calculated FRiP values for each cell barcode, providing an indicator of signal-to-noise ratio in single-cell epigenomic data.

### 5.4 CAR Insertion Site Coverage Heatmap

The heatmap shows per-cell read coverage across CAR vector insertion sites (e.g., EF1a, PGK1, TRAC). Each row represents a cell barcode, and each column represents a CAR target site. This allows visualization of which cells are likely to carry vector insertions and where signal enrichment occurs, helping to verify construct delivery or identify mosaicism in editing (Torres et al. 2022).

### 5.5 Top CAR Sites by Total Coverage

This table summarizes the top 10 CAR insertion sites ranked by the total number of reads mapped to each site across all cells. It helps identify the most active or consistently covered transgene targets, which is critical in validating vector design and insertion stability (Frangieh et al. 2021).

# 6 Top CAR Insertion Site Read Counts

| Site                | Total Reads |
|---------------------|-------------|
| B2M_KO_CAR_Target   | 91          |
| TRAC_CAR_Eyquem2017 | 53          |
| RAB11A_CAR_Target   | 24          |
| CCR5_CAR_Target     | 18          |
| AAVS1_CAR_Common    | 4           |

Table 2: Top CAR Insertion Site Read Counts: Read counts mapped to the top CAR (Chimeric Antigen Receptor) insertion sites. Each row represents a genomic target site with the corresponding number of supporting reads, indicating the relative abundance of CAR integrations across different loci.

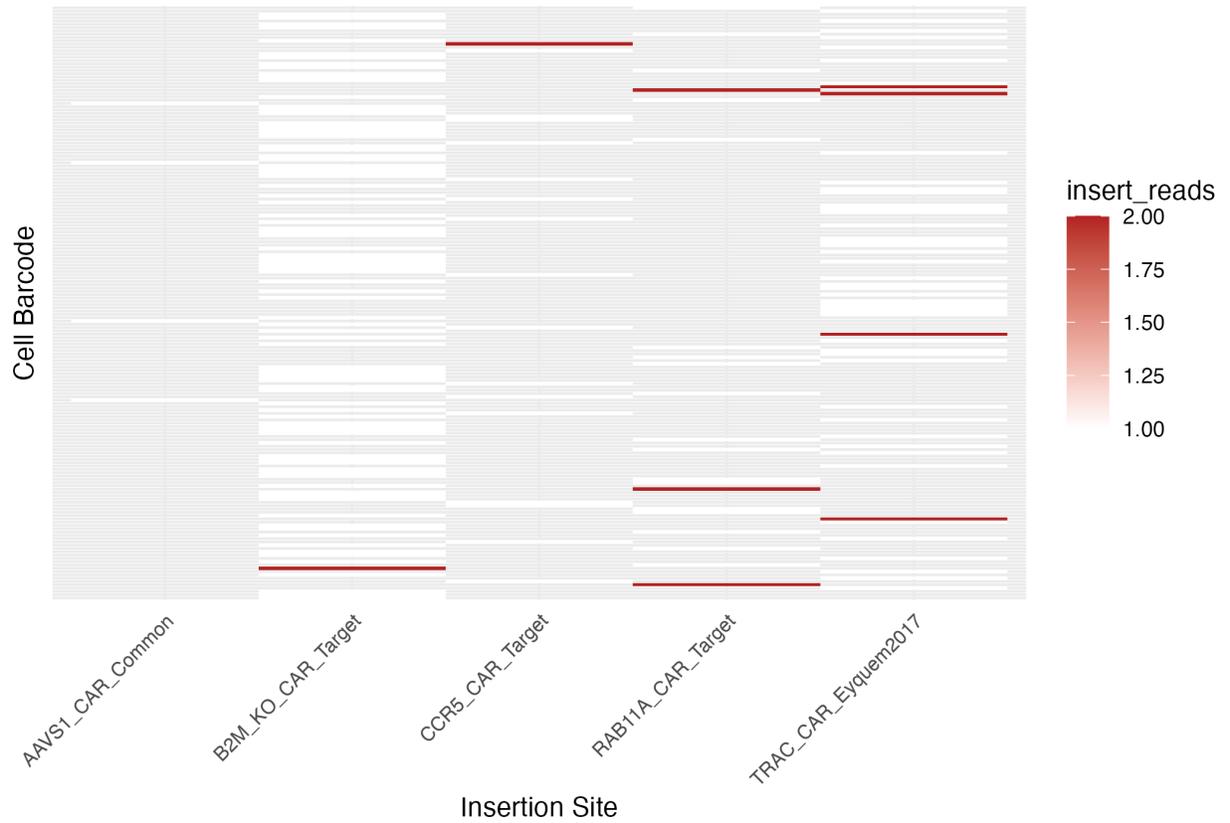

Figure 2: CAR Insertion Heatmap: Heatmap showing CAR (Chimeric Antigen Receptor) insertion site coverage across individual cells. Each row corresponds to a target site, and each column represents a single cell, with color intensity indicating the number of reads supporting integration. This highlights the distribution and specificity of CAR insertions in the dataset.

# 7 Repository and Installation

The source code is hosted on GitHub:
https://github.com/biosciences/CARTEpigenoQC

# 8 Acknowledgements

The author thanks collaborators at the University of Sydney for insights into CAR-T clinical pipelines, and for pipeline development guidance.